\documentclass[review]{elsarticle}

\usepackage{epstopdf}
\journal{Journal of \LaTeX\ Templates}

\begin{document}

\begin{frontmatter}

\title{Prediction of MoO$_{\rm 2}$ as high capacity electrode material for (Na, K, Ca)$-$ion batteries\tnoteref{mytitlenote}}

\author{Yong-Chao Rao, Song Yu, Xiao Gu}

\author{ Xiang-Mei Duan$^{\ast}$}
\address{Department of Physics, Faculty of Science, Ningbo University, Ningbo-315211, P.R. China}
\cortext[mycorrespondingauthor]{Corresponding author}
\ead{duanxiangmei@nbu.edu.cn}

\begin{abstract}
Na, K and Ca$-$ion battery electrode materials with appropriate electrochemical properties are desirable candidates for replacing lithium$-$ion batteries (LIBs) because of their natural richness and low cost.
Recently, MoO$_{\rm 2}$ has been reported as the anode material in LIBs, but so far not received attention in Na and other ion batteries. In this paper, the behaviors of Na, K, and Ca on MoO$_{\rm 2}$ are investigated by first$-$principles calculations. These metal atoms strongly absorb on the hexagonal center of MoO$_{\rm 2}$
 and the adsorption results in semiconducting$-$metallic transition. The low diffusion barrier, 0.13, 0.08, 0.22~eV for Na, K, Ca, respectively, leads to an ultrahigh diffusivity.  Importantly, the maximum metal$-$storage phases of MoO$_{\rm 2}$ monolayer correspond to Na$_{\rm 4}$MoO$_{\rm 2}$, K$_{\rm 2.5}$MoO$_{\rm 2}$ and Ca$_{\rm 3}$MoO$_{\rm 2}$, with considerable theoretical specific capacities of 837, 523 and 1256 mAh g$^{\rm -1}$.
 The electrode materials exhibit moderate average voltage of 0.30, 0.75, 0.35 V, respectively.
 Our findings suggest that MoO$_{\rm 2}$ monolayer can be utilized as a promising anode material with high capacities and high rate performance for next generation ion batteries.
\end{abstract}

\begin{keyword}
\texttt MoO$_{\rm 2}$ monolayer; Ion$-$battery; First principles calculations
\end{keyword}

\end{frontmatter}

\section{Introduction}
With the tremendous and rapid development of portable electronic devices and electric vehicles, rechargeable batteries with large capacity, high rate capability and good cycle performance have attracted great attention both in basic research and industry applications.\cite{Dunn2011Electrical} As one of the most important energy storage devices, lithium$-$ion batteries (LIBs) are widely used in the commercial society.\cite{Manthiram2013In} However, issues such as cost, safety, energy storage density, charge/discharge rates, and lifetime will continue to spur the research for new batteries beyond LIBs.\cite{Goodenough2013The, Tarascon2010Is} Room temperature sodium$-$ion batteries (SIBs) have gained increasing attention due to the cheaper cost and relative abundance of sodium.\cite{Kim2012Electrode, Palomares2012Na} Calcium$-$ion batteries (CIBs) offer low price, chemical safety and lighter mass$-$to$-$charge ratio.\cite{Hayashi2003Electrochemical} Moreover, potassium is indeed a competitive metal with less complicated interfacial reaction \cite{Moshkovich2001Investigation} and higher ionic conductivity \cite{Eftekhari2016Potassium} in solution.

The battery$-$performance characteristic strongly depends on the electrochemical properties of electrode materials. Great challenge of advanced metal$-$ion batteries is to find better electrode substrates. Currently, a wide range of compounds have been proposed for SIB anodes, and much progress has been made .\cite{Clynen2013ChemInform, Kim2012Electrode, Park2013Anomalous} However, capacity for SIBs are still limited owing to the relatively large ionic radius of Na$-$ion (1.02 \AA) compared to Li$-$ion (0.76 \AA).
The commercially used anode material in LIBs, graphite, is not suitable for Na$-$ion batteries for the low capacity and poor rate capability.\cite{Winter2010Insertion} With the high specific surface area, 2D materials provide variable choice in designing high energy and high motility devices.\cite{Yoo2008Large, Yu2018Net}  Much research efforts turn to the development of SIBs, KIBs (potassium ion$-$battery) and CIBs based on 2D materials, such as graphene\cite{Malyi2015A} and its dopant,\cite{Datta2014Defective, Gong2017Boron} transition metal oxides,\cite{Caballero2002Synthesis, Liu2011Electrochemical} TMDs,\cite{Shuai2016Density, Yang2015Two} silicene\cite{Zhu2016Silicene} and other van der Waals$-$bonded layered materials.\cite{Clynen2013ChemInform, David2014MoS2} Recently, MoO$_{\rm 2}$ sub$-$micro sheet was synthesized by using a novel chemical vapour deposition method.\cite{Liu2011Preparation} The dynamical and thermal stabilities MoO$_{\rm 2}$ monolayer have been confirmed via {\it ab initio} molecular dynamics simulation and phonon dispersion spectrum. Moreover, Li$-$binding on MoO$_{\rm 2}$ sheet possesses superb conductivity, lower diffusion barrier (75~meV), and astonishing theoretical storage capacity (2513~mAh g$^{-1}$).\cite{Zhou2017MoO2}

Due to the electrochemical similarity between LIBs and SIBs, KIBs and CIBs, an interesting question arises: Can the superior electrochemical property of the MoO$_{\rm 2}$ sheet, demonstrated in LIBs, extend into Na, K, Ca$-$ion cells? In this work, DFT calculations are employed to systematically investigate the electronic, metal adsorption and storage behaviors of MoO$_{\rm 2}$ monolayer. Our results indicate that Na, K and Ca atoms can strongly absorb on the surface of MoO$_{\rm 2}$, and the metallic characteristics they presented, are crucial for the anode materials. The diffusion barrier for Na, K, Ca on MoO$_{\rm 2}$ is only 0.13, 0.08, 0.23~eV, respectively, implying good diffusion mobility during the charging/discharging process. Furthermore, the theoretical specific capacities are as high as 837, 523, 1256 mAh g$^{\rm -1}$ for Na, K, Ca$-$ion batteries. Most especially, the capacity for Na is 4.5 times higher than that of the oxidized graphite electrode (184~mAh g$^{\rm -1}$).\cite{Yang2014Expanded} Therefore, as electrode materials for Na, K, Ca$-$ion batteries, the MoO$_{\rm 2}$ sheet possesses great potential and outstanding performance.

\section{Computational details}

DFT calculations were performed by using the Vienna ab initio Simulation Package (VASP). The exchange$-$correction interaction were treated within the generalized gradient approximation (GGA) in the form of the Perdew$-$Burke$-$Ernzerhof (PBE) functional.\cite{Perdew1996Generalized} The electron wave functions were expanded using plane waves with a cutoff energy of 500~eV, and the convergence criteria for the residual force and energy on each atom during structure relaxation was set to 10$^{\rm -2}$~eV/{\AA} and 10$^{\rm -5}$~eV, respectively.
$4\times4\times1$ supercells were built and complemented by vacuum layers of 20 {\AA} thickness in the out$-$of$-$plane direction to avoid interaction between periodic images. The $k-$point sampling of $3\times3\times1$ and $7\times7\times1$ were chosen for geometry optimization and electronic structure computation, respectively. For the adatom diffusion on MoO$_{\rm 2}$ sheet, the climbing image nudge elastic band method was used to perform minimum energy path profiling and figure out the diffusion barrier.\cite{Henkelman2000Improved}

The strength between the metal atoms and the substrate was described by the average adsorption energy, $E_{\rm ads}$, which is defined as
\begin{equation}
E_{\rm ads} = (E_{{{\rm A}_{x}\rm{MoO}_{\rm 2}}} - E_{\rm {MoO_{\rm 2}}} - x\mu_{\rm A})/x
\end{equation}
where $x$ is the total number of adsorbed metal atoms in the primitive cell and $E_{{{\rm A}_{x}\rm{MoO}_{\rm 2}}}$, $E_{\rm {MoO_{\rm 2}}}$, $\mu_{\rm A}$ is total energies of the metal ion adsorbed on the system, pristine MoO$_{\rm 2}$ monolayer, and the energy per atom in the bulk metal, respectively.

The average voltage of A$_{x}$MoO$_{\rm 2}$ (A $=$ Na, K, and Ca) in the range of $x_{\rm 1} \leq x \leq x_{\rm 2}$ is given as
\begin{equation}
V_{\rm ave} \approx [E_{{{\rm A}_{x_{\rm 1}}\rm{MoO}_{\rm 2}}} - E_{{{\rm A}_{x_{\rm 2}}\rm{MoO}_{\rm 2}}} + (x_{\rm 2} - x_{\rm 1})\mu_{\rm A}]
/ (x_{\rm 2} - x_{\rm 1})e
\end{equation}

According to the Faraday equation, the maximum specific capacity, $C_{\rm A}$, is obtained by\cite{Datta2014Defective}
\begin{equation}
C_{\rm A} = x_{\rm max}F/M
\end{equation}
Where $x_{\rm max}$ represents maximum number of A in the primitive cell, $F$ is the Faraday constant with the value of 26806~mAh~mol$^{-1}$, and $M$ is the mass of MoO$_{\rm 2}$ in g~mol$^{-1}$.

\section{Results and discussion}
\subsection{Adsorption of Na, K, Ca on MoO$_{\rm 2}$ surface}

As the common transition$-$metal dichalcohenides (TMDs), MoO$_{\rm 2}$ monolayer is composed of a triple layer (sandwich structure), where atoms are stacked in a sequence of O$-$Mo$-$O with thickness of  2.47~\AA, the optimized lattice parameter is found to be $a = b = 2.83$~\AA, which are in good agreement with the previous calculation values of $2.82$~\AA. \cite{Rasmussen2015Computational}

To systematically study metal atoms intercalation on MoO$_{\rm 2}$ monolayer, we first examine four possible high$-$symmetric adsorption sites (H site: hexagonal center site; B site: over the center of the Mo$-$O bond; T site: the top of a O atom; T$'$ site: the top of a Mo atom) for isolated Na, K, Ca atom, as plotted in Fig. 1 (a). The adsorption energies ($E_{\rm ads}$) of a single Na, K, and Ca atom adsorbed on MoO$_{\rm 2}$ are illustrated in Fig. 1 (c). In general, the K$-$adsorbed configurations are the most stable, while the strength of adsorption of one Ca atom on MoO$_{\rm 2}$ is the weakest with the smallest $E_{\rm ads}$. The negative adsorption energies for H, B and T$'$ configurations indicate the Na, K, and Ca storage is exothermic and spontaneous, which is foundational and favorable for the application. H site is the most stable configuration with the largest adsorption energy of $-$1.46, $-$2.01, $-$1.16 eV for Na, K and Ca, respectively. We note that a single Na atom adsorbed on the MoO$_{\rm 2}$ monolayer is more stable, when compared with other 2D materials, such as boron$-$doped graphene ($-$0.79~eV)\cite{Ling2014Boron} and single$-$layer MoS$_{\rm 2}$ ($-$1.23~eV).\cite{Su2014Ab} High adsorption strength ensures that metal atoms do not form clusters, which is beneficial to keep good cycling during the charging/discharging process of batteries.
More particularly, the adsorption energies on T$'$ site ($-$1.42, $-$1.99, $-$1.07 eV for Na, K and Ca, respectively) are slightly smaller that that of H site, which implies that the metal atoms may move from the H site to the T$'$ site with lower energy barrier.

Conductivity is a key factor in determining the electrochemical performance of an electrode.
Pristine MoO$_{\rm 2}$ is an indirect semiconductor with a band gap of 0.92~eV, which is consistent with previous theoretical work of 0.91~eV,\cite{Rasmussen2015Computational} implying poor electrical conductivity.
The valence band maximum (VBM) of MoO$_{\rm 2}$ is contributed by O$-2p$ orbitals and its conduction band minimum (CBM) is dominated by Mo$-4d$ orbitals.
The total density of states (TDOS) and projected density of states (PDOS) for the adsorption of single Na, K, Ca, respectively, are shown in Fig. 2 (a), (b) and (c).
It can be seen that three adsorbed systems exhibit metallic characteristic, the CBM of each system is still dominated by Mo 4$d$ orbital,
and the $s$ orbital of metal atoms is empty. To obtain insights into the nature of metals$-$MoO$_{\rm 2}$ binding, we refer to the charge density differences of (110) section defined as $\Delta\rho=\rho_{\rm metal/MoO_2}-\rho_{\rm metal}-\rho_{\rm MoO_2}$, which are shown in Fig. 3. Clearly, electrons accumulate around oxygen atoms and their densities surrounding metal atom decrease, suggesting ionic character of metals$-$MoO$_{\rm 2}$ binding. By using Bader charge analysis, the absorbed Na, K and Ca atoms act as electron donor, and transfer 0.9, 0.93 and 1.61~$e$ to the underlying MoO$_{\rm 2}$, respectively.

The spin$-$dependent GGA plus Hubbard U (GGA$+$U) method was employed as a test. Comparing the $E_{\rm ads}$ and DOS with the results obtained from the common GGA method, the difference is negligible, implying the weak onsite Coulomb repulsion between the 4$d$ electrons of molybdenum. Moreover, the systems with/without metals absorption show non$-$magnetic behavior.

\subsection{Diffusion of metal adatoms on MoO$_{\rm 2}$ surface}

Another factor that plays a central role in the performance of electrode materials is the diffusion rate of ions, which contributes directly to the charging/discharging loading rates accessible by a battery.
We investigate the diffusion pathways and energy barriers of Na, K and Ca on MoO$_{\rm 2}$ surface. Only the diffusion path between two neighboring, most preferred binding sites are taken into consideration. Due to the symmetry, there are three representative pathways for metal atom migration between two adjacent H sites: (a) through the T site (H$-$T$-$H), (b) across the Mo$-$O bond (H$-$B$-$H), and (c) through the T$'$ site.
As we mentioned above, the almost equivalent adsorption energy at H and T$'$ sites motivates us to only consider the pathway (c). The optimized route is shown in Fig. 4(a).
The corresponding energy profiles of Na, K and Ca along pathway on the MoO$_{\rm 2}$ surface are depicted in Fig. 4(b). The barrier for Na is 0.13~eV, implying very fast Na diffusion kinetics, which would lead MoO$_{\rm 2}$ to be an effective anode electrode for Na$-$ion batteries. The value of the barrier is comparable with other 2D materials such as MoS$_{\rm 2}$ (0.11~eV),\cite{Su2014Ab} graphene (0.13~eV)\cite{Datta2014Defective} and boron$-$doped graphene
(0.16$-$0.22 ~eV).\cite{Ling2014Boron} Particularly, the barrier in K$-$adsorbed system is as low as 0.08~eV, smaller than corresponding Na system, which can be ascribed to the larger vertical distance (2.16~\AA) between K and substrate, while Na$-$substrate distance is 1.68~\AA. Similar trends have been observed in VS$_{\rm 2}$\cite{Momeni2016Maleic} and MoN$_{\rm 2}$.\cite{Zhang2016Theoretical} For the case of Ca, the relative large mass results in a short cation$-$substrate distance of 1.50~\AA, the diffusion barrier of Ca is 0.22~eV. Therefore, we conclude that MoO$_{\rm 2}$ monolayer have the ability to rapidly charge/discharge Na and K ions.

\subsection{Average open$-$circuit voltage and metals storage capacity of the MoO$_{\rm 2}$ monolayer}

From the practical point of view, the average open$-$circuit voltage and metal storage capacity are significant features for the electrode materials.
Due to the high symmetry of MoO$_{\rm 2}$ structure, we consider all the adsorption with double sides to obtain the higher capacity.
The typical Na$_x$MoO$_{\rm 2}$ configurations are illustrated in Fig. 5. To avoid the inherently strong coulomb repulsion as metals absorb on the top of the MoO$_{\rm 2}$ surface, the adsorption structures appear to be mixed configurations. As shown in Fig. 5 (a), four Na atoms absorb on the H site due to its relatively larger $E_{\rm ads}$. The additional four Na atoms tend to adsorb on the next nearest neighbor T$'$ site, as shown in Fig. 5(b), to reduce the repulsion as possible. As the concentration increases, the physical attraction between the further far metal atoms and MoO$_{\rm 2}$ becomes weaker, so the adsorption prefers to proceed layer by layer, with each layer containing eight metal atoms [see Na$_{\rm 3}$MoO$_{\rm 2}$ in Fig. 5 (c)].

 In Fig. 6, we present the OCVs as function of the concentration by varying $x$ in the A$_x$MoO$_{\rm 2}$ system.
 With the increase of $x$, the OCVs decrease gradually because the distances between adjacent metal atoms become shorter and their repulsive interaction is enhanced. Once the OCV is zero, no more atom can be adsorbed, and the maximum $x$ in A$_x$MoO$_{\rm 2}$ is obtained. Encouragingly, the large initial OCVs of 1.27, 1.79, and 1.16 V for Na, K, and Ca systems, respectively, indicate the potential application of MoO$_{\rm 2}$ in ion$-$batteries.
 For the Na$_x$MoO$_{\rm 2}$ system, the electrode potential varies in the range of 1.27$-$0.09 V. The $V_{\rm ave}$, 0.30 V, is in the desirable voltage range for a sodium anode of 0.0$-$1.0 V. At $x = 4$, the corresponding maximum specific capacity of MoO$_{\rm 2}$ for SIBs is 837 mAh g$^{\rm -1}$, which is larger than that of existing SIB anode materials (including hard carbon,\cite{Dahbi2014Negative, Kundu2015ChemInform} Na$_{\rm 4}$Ti$_{\rm 5}$O$_{\rm 12}$,\cite{Naeyaert2015ChemInform} Na$_{\rm 2}$Ti$_{\rm 3}$O$_{\rm 7}$\cite{Senguttuvan2011ChemInform}) and is even larger than some alloy$-$based materials.\cite{Komaba2012Redox, Qian2012High}
 Compared with the capacities of 2D materials, such as MoS$_{\rm 2}$/graphene of ~338 mAh g$^{-1}$,\cite{David2014MoS2} VS$_{\rm 2}$ of 232.91 and 116.45 mAh g$^{\rm -1}$ for 1H and 1T phases, respectively\cite{Putungan2016Metallic} and GeS nano$-$sheet of 512 mAh g$^{\rm -1}$,\cite{Li2016Germanium} the capacity of Na$_{\rm 4}$MoO$_{\rm 2}$  is also competitive.
 Therefore MoO$_{\rm 2}$ monolayer should be a promising candidate for an electrode material in SIBs.
For the case of K and Ca adsorbed MoO$_{\rm 2}$, electrode potentials are in the range of 0.02$-$1.79~V and 0.25$-$1.16~V, with the $V_{\rm ave}$ of 0.75 and 0.35~V, respectively. Due to the large ionic radius of K and Ca, the maximum $x$ is estimated to be 2.5 and 3, and the corresponding capacity of KIB and CIB is 523 and 1256~mAh g$^{\rm -1}$.

When evaluating cycling stability, the degree of structural deformation would cause substantial loss of capacity in several cycles. After the absorption of maximum Na, K, and Ca, the thickness of MoO$_{\rm 2}$ is 2.50, 2.48, 2.52 {\AA} respectively, slightly stretched from the original thickness of 2.47 {\AA}.
The bond length of Mo$-$O changes negligibly.
During the process of ions intercalation/deintercalation, the structure remains intact to ensure good cycle performance and a smooth surface that will undoubtedly minimizes diffusion resistance during the metal ion transport.
So far, the MoO$_{\rm 2}$ anode for KIBs is predicted to be a perfect material, which simultaneously possesses a large specific capacity (523~mAh g$^{\rm -1}$), excellent stability and conductivity [see Fig. 2(b)], and a modest open circuit voltage (0.75 V).

\section{Conclusion}
In summary, we investigate the properties of 2D MoO$_{\rm 2}$ monolayer as potential anodes for batteries beyond LIBs.
With large adsorption energies, Na, K or Ca atom is preferentially absorbed over the center of the honeycomb lattice, and the charge transfer from metal atoms to the substrate results in a transition of semicondutive MoO$_{\rm 2}$ to metallic A$_{x}$MoO$_{\rm 2}$.
The diffusion barriers of metal atoms on the surface of MoO$_{\rm 2}$ are relatively low, especially for Na and K, which are only 0.13 and 0.08~eV.
The theoretical specific capacity of Na$_{\rm 4}$MoO$_{\rm 2}$, K$_{\rm 2.5}$MoO$_{\rm 2}$ and Ca$_{\rm 3}$MoO$_{\rm 2}$ are as high as 837 mAh g$^{\rm -1}$, 523 mAh g$^{\rm -1}$ and 1256 mAh g$^{\rm -1}$, respectively. When the MoO$_{\rm 2}$ anodes are used for SIBs, KIBs and CIBs, the open voltage is 0.30~V, 0.75~V and 0.35~V, respectively.
Moreover, MoO$_{\rm 2}$ exhibits high mechanical stability and integrity upon metal insertion.
Our results give insights for further experimental work in exploring the possibility of MoO$_{\rm 2}$ for Na, K and Ca$-$ion batteries.

\section*{Acknowledgments}

This research is supported by the Natural Science Foundation of China (grant No. 11574167), the New Century 151 Talents Project of Zhejiang Province and the KC Wong Magna Foundation in Ningbo University.

\section*{Conflict of interest}
The authors declare they have no conflict of interest

\section*{References}

\bibliography{references}
\newpage
\begin{figure}[h]
 \centering
 \includegraphics[width=10cm]{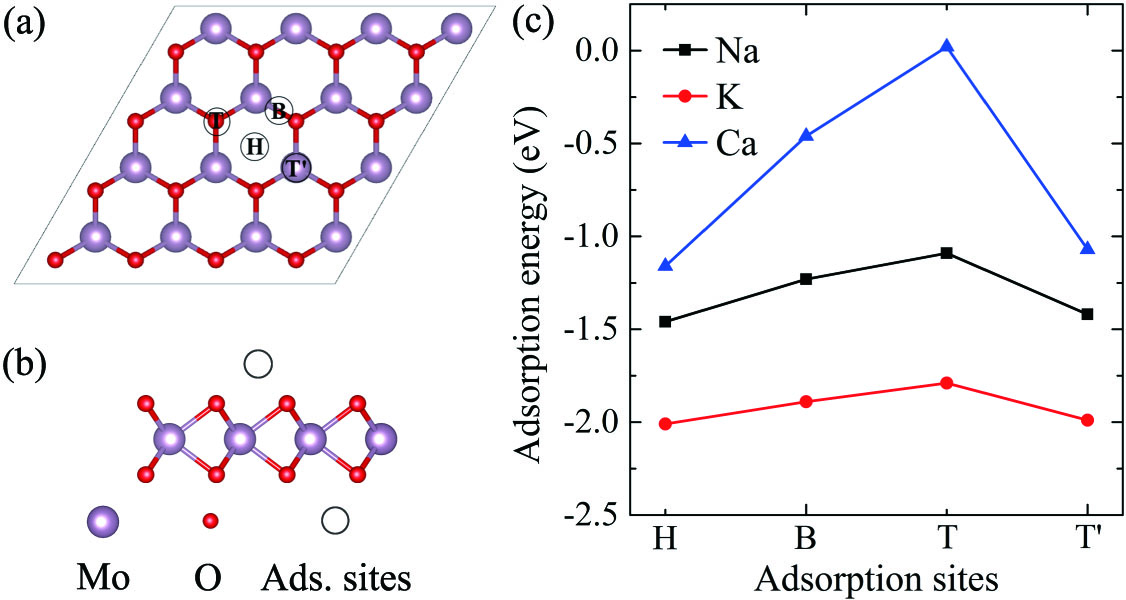}
 \caption{(a) Top view for the adsorption of metal atom at different sites on the MoO$_{\rm 2}$ sheet. (b) Side view of adsorbate on the H site. (c) The adsorption energy for Na, K, Ca based on the four distinct binding sites on MoO$_{\rm 2}$.}
\end{figure}

\begin{figure}[h]
  \centering
  \includegraphics[width=8cm]{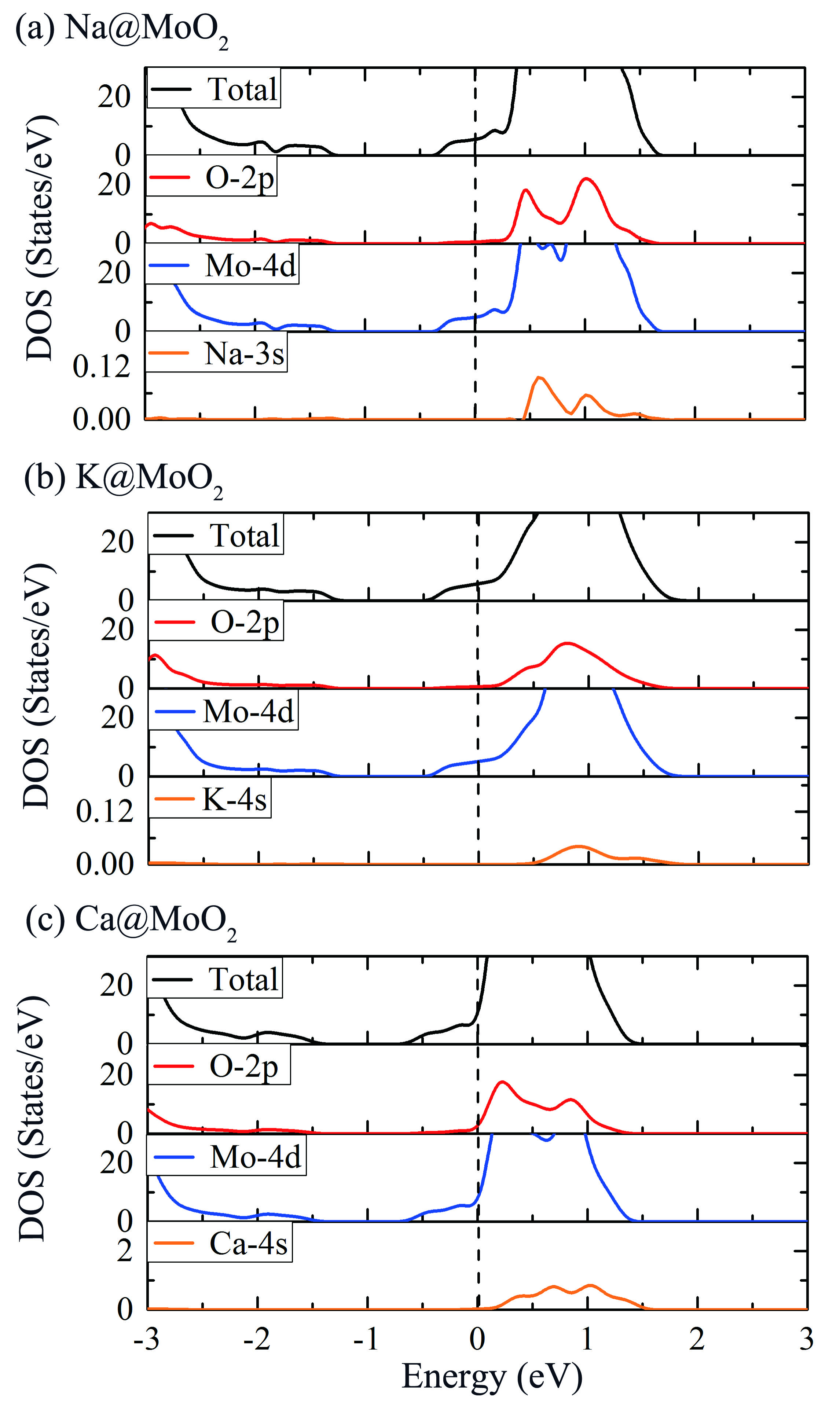}
  \caption{Total and projected density of states for (a) Na@MoO$_{\rm 2}$, (b) K@MoO$_{\rm 2}$, and (c) Ca@MoO$_{\rm 2}$, respectively. The horizontal dashed lines represent the Fermi level. }
\end{figure}

\begin{figure}[h]
  \centering
  \includegraphics[width=9cm]{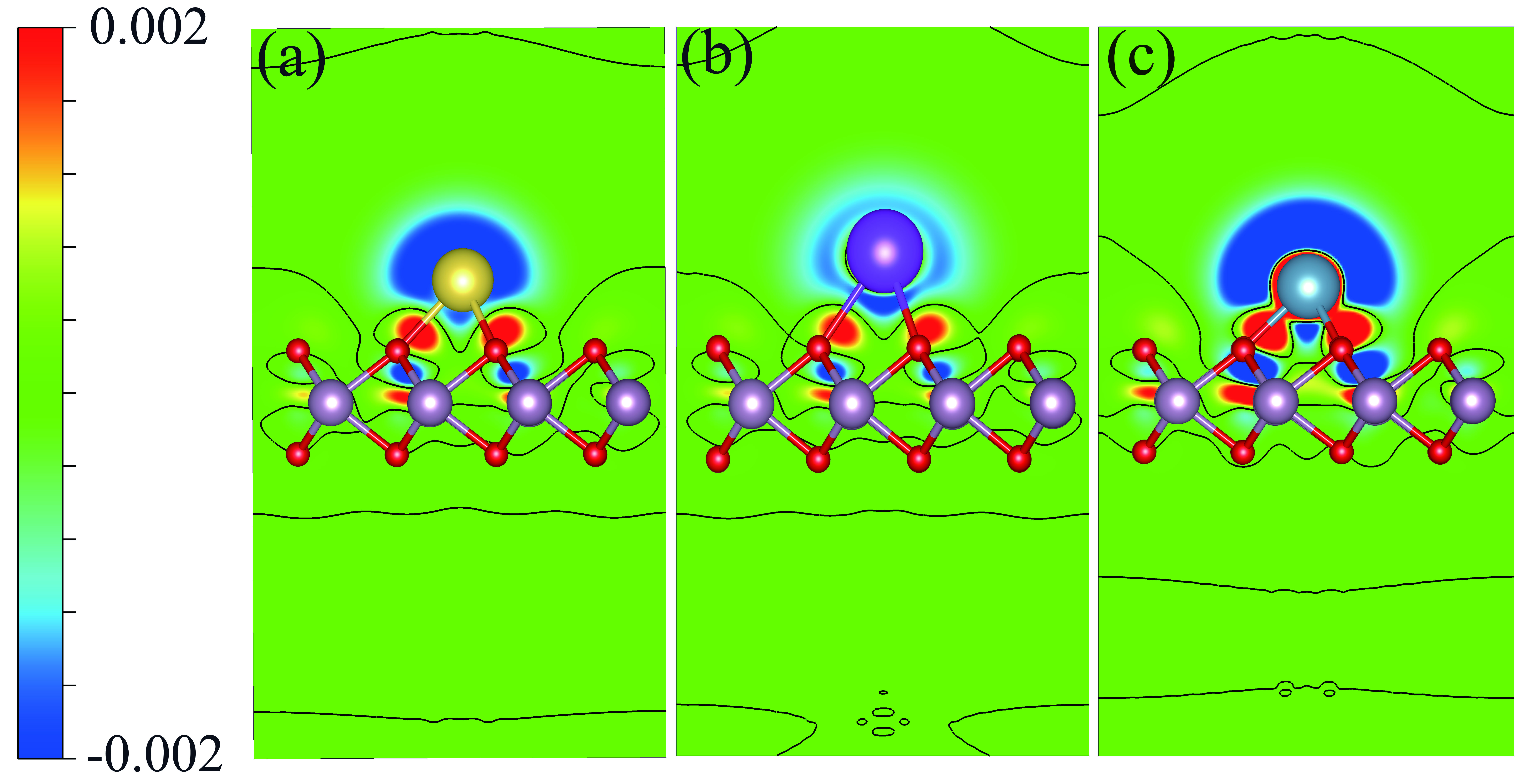}
  \caption{Charge density difference of the (110) section for single adatom (a) Na, (b) K and (c) Ca (being adsorbed on the H site) on the MoO$_{\rm 2}$. Red and blue colors indicate the electrons accumulation and depletion, respectively. The color scale is in the units of e/\AA$^{\rm 3}$. }
\end{figure}

\begin{figure}[h]
  \centering
  \includegraphics[width=9cm]{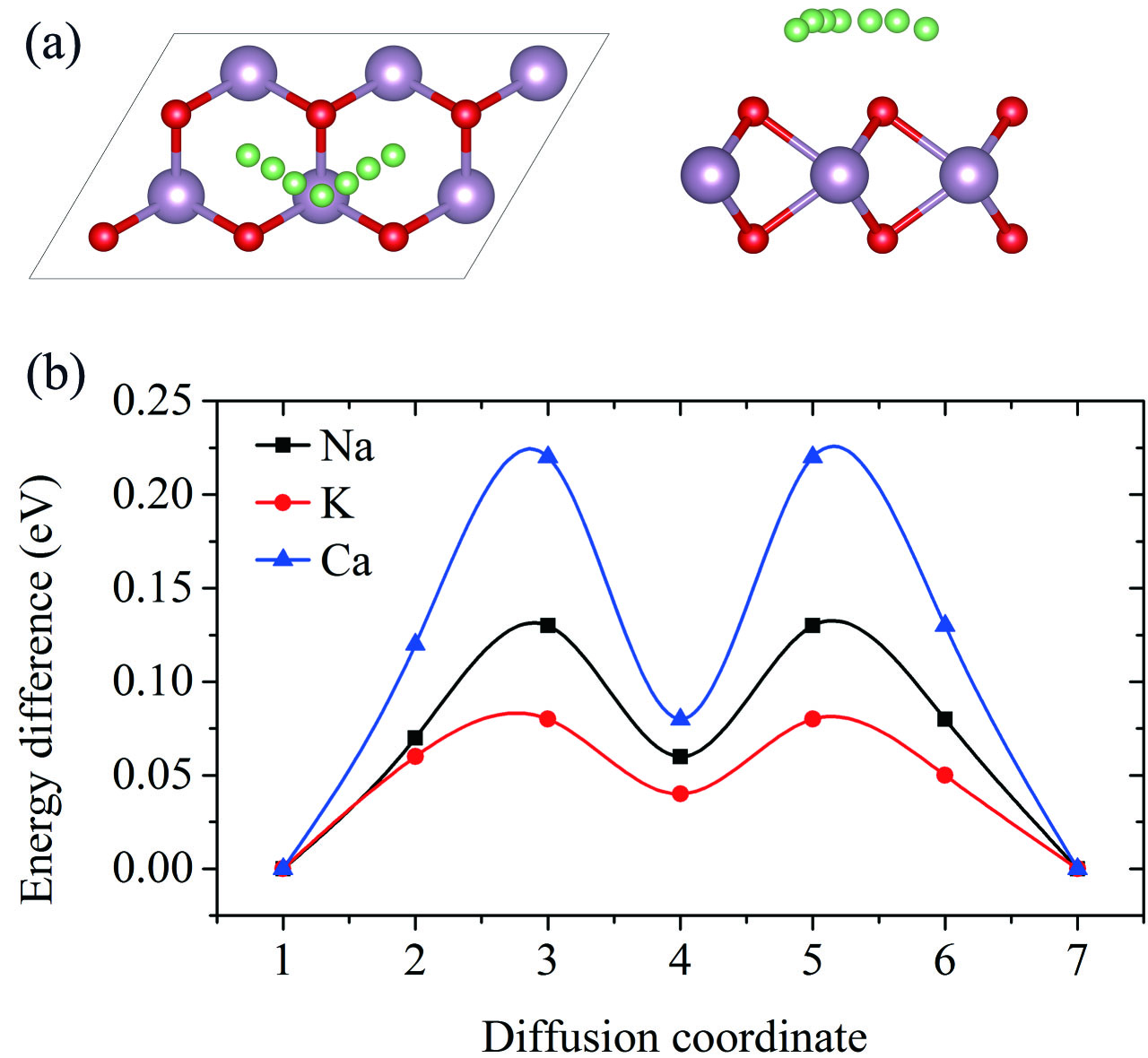}
  \caption{(a) the top and side views of the optimized migration pathways and the corresponding diffusion barrier profiles of (b) Na (black), K (red), Ca (blue) on the MoO$_{\rm 2}$ sheet, respectively. }
\end{figure}

\begin{figure}[h]
  \centering
  \includegraphics[width=16cm]{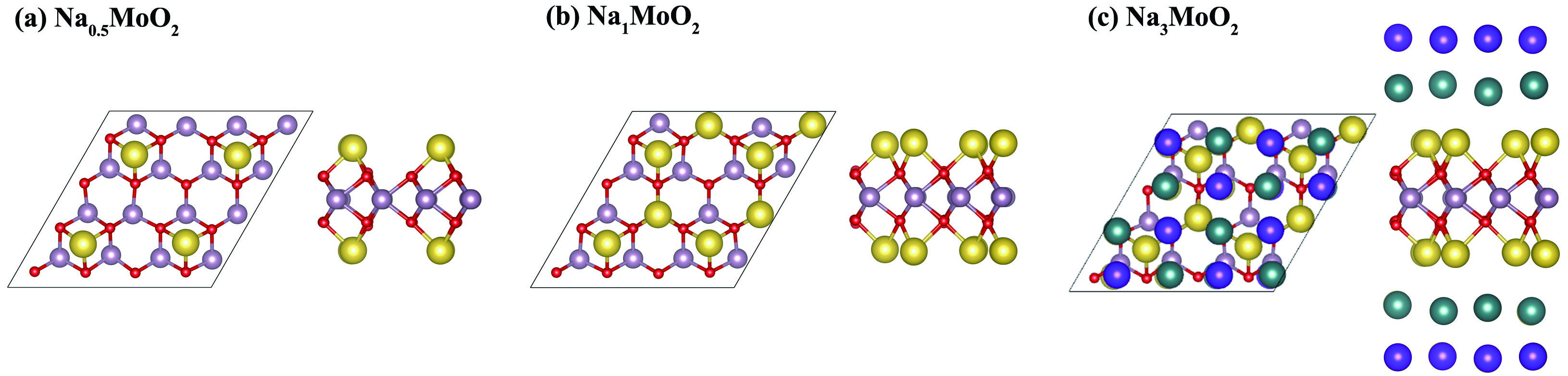}
  \caption{Top and side views of the optimized structures of (a) Na$_{\rm 0.25}$MoO$_{\rm 2}$, (b) Na$_{\rm 1}$MoO$_{\rm 2}$, and (c) Na$_{\rm 3}$MoO$_{\rm 2}$. The color of yellow, green and violet represent the first, second and third Na layer, respectively.}
\end{figure}

\begin{figure}[h]
  \centering
  \includegraphics[width=9cm]{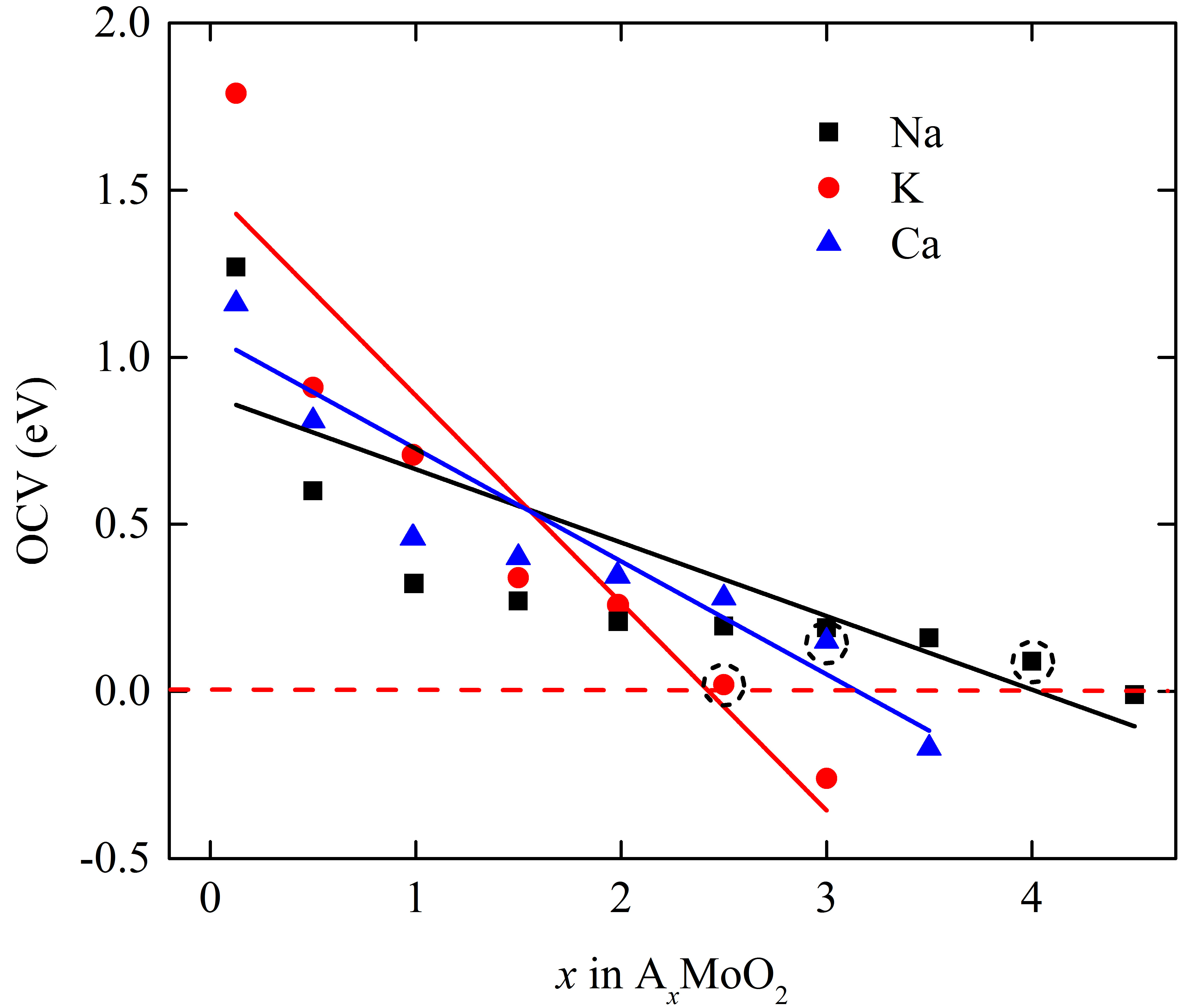}
  \caption{The variation of OCVs with the increasing concentration of metal atoms on the MoO$_{\rm 2}$ surface. The dashed open circles correspond to the maximum $x$ in the compounds}
\end{figure}
\end{document}